\documentclass[aps,pre,floats,twocolumn,twoside,floatfix,superscriptaddress]{revtex4}

\usepackage{graphicx}
\usepackage{color}
\usepackage{amsmath,amssymb}
\usepackage{psfrag}
\usepackage[utf8]{inputenc}

\begin{document}

\title{A branching random-walk model of disease outbreaks and the percolation backbone}

\author{Paulo Murilo C. de Oliveira}
\email{pmco@if.uff.br}
\affiliation{Instituto de Física, Universidade Federal Fluminense, Av. Litorânea s/n, 24210-340 Boa Viagem, Niterói, RJ, Brazil}
\affiliation{Instituto Nacional de Ciência e Tecnologia - Sistemas Complexos, Rio de Janeiro RJ, Brazil}

\author{Daniel A. Stariolo}
\email{danielstariolo@id.uff.br}
\affiliation{Instituto de Física, Universidade Federal Fluminense, Av. Litorânea s/n, 24210-340 Boa Viagem, Niterói, RJ, Brazil}
\affiliation{Instituto Nacional de Ciência e Tecnologia - Sistemas Complexos, Rio de Janeiro RJ, Brazil}

\author{Jeferson J. Arenzon}
\email{arenzon@if.ufrgs.br}
\affiliation{Instituto de F\'\i sica, Universidade Federal do Rio Grande do Sul, CP 15051, 91501-970 Porto Alegre RS, Brazil}
\affiliation{Instituto Nacional de Ciência e Tecnologia - Sistemas Complexos, Rio de Janeiro RJ, Brazil}

\date{\today}

\begin{abstract}
The size and shape of the region affected by an outbreak is relevant to understand the dynamics of a disease and help to organize future actions to mitigate similar  events. 
A simple extension of the SIR model is considered, where agents diffuse on a regular lattice and the disease may be transmitted when an infected and a susceptible agents are nearest neighbors.  
We study the geometric properties of both the connected cluster of sites visited by infected agents (outbreak cluster) and the set of clusters with sites that have not been visited.
By changing the density of agents, our results show that there is a mixed-order (hybrid) transition where the region affected by the disease is finite in one phase but percolates through the system beyond the threshold.
Moreover, the  outbreak cluster seems to have the same exponents of the backbone of the critical cluster of the ordinary percolation while the clusters with unvisited sites have a size distribution with a Fisher exponent $\tau<2$. 
\end{abstract}

\maketitle

\section{Introduction}

An epidemic outbreak may be controlled once herd immunity develops because most of the population has been either vaccinated or infected by the contagious agent and recovered. 
Amidst those individuals that no longer can get infected, groups of yet susceptible ones may remain. 
The sizes of these groups and the distances and connecting routes between them will determine the consequences of a future outbreak. 
Thus, along with the temporal course of an epidemic, for future prevention and mitigation of similar events it is also important to model and characterize both the spatial extent~\cite{DuMaRoZo13,Bordeu19} and the geometric properties of the regions that had been affected. 
Moreover, the interest in spreading processes goes beyond the propagation of diseases, since epidemiological models consider general mechanisms that have been applied, for example, in studies of how opinions, gossips, or fake news propagate through the complex networks of social contact~\cite{CaFoLo09,Baronchelli18,AmAr18,DaAmAr20}. 
 
A cornerstone of many theoretical studies in epidemiology is the SIR model and its many variants (see Refs.~\cite{PaCaMiVe15,SuJuJiWaWa16,Brauer17,ArRoMo18,Mata21} for recent reviews). 
Infectious agents (I) may either transmit the disease, with a given rate, to susceptible ones (S) through direct contact or, spontaneously, get removed (R) from the process by dying or recovering. 
This compartmental model has an absorbing state where the infected individuals get extinct and the disease can no longer propagate. 
Depending on the parameters, there is a continuous transition between a phase where the outbreak is controlled fast, leaving only a finite number of agents that have been infected, and another phase where most of the agents became infected and, once recovered, occupy a macroscopic fraction of the lattice. 
The nature of this transition, belonging to the percolation universality class, has been largely studied~\cite{CaGr85,ToZi10,SoToZi11}. 
Several modifications have been introduced in the original SIR model (further compartments, non-permanent immunity, different networks and lattices, disorder, vaccination, etc) and both the asymptotic and dynamical properties of the contagion process in these scenarios have been studied, along with the possibility of changing the universality class of the transition. 
In some cases, the transition changes from continuous to discontinuous, or even to a hybrid (or mixed order) one, where the discontinuity in the order parameter is accompanied by critical fluctuations~\cite{Thouless69,Cardy81}.  
Such hybrid transitions have been observed in models with cooperative spreading where multiple strains (or multiple contagion steps) are involved~\cite{DoWa04,BiPaGr12,CaChGhGr15,BaLiScZi15,JaSt16,LeChKeKa17}. 

Analogously to the SIR model, the system we consider in this paper is a stochastic branching process, where the agents are random walkers with volume exclusion~\cite{BoCh92,Zhang2005,SiFe15} and the infection proceeds through close encounters between infected agents and susceptible ones. 
Once the outbreak is over, we study the spatial extent of the infected region, i.e., the asymptotic set of all sites that have been visited by an infected agent before recovery~\cite{DuMaRoZo13,Bordeu19}. 
Depending on the parameters of the model, there is a phase transition connected with the size of this region. 
The main purpose of this paper is to characterize this transition and determine which is the universality class this model belongs to. 
Besides studying the cluster of sites visited by infected agents, it is also interesting to study those that were not. As the number of agents increase, facilitating the propagation of the disease, the set of unvisited sites appears fragmented in several independent regions. The geometric properties of these clusters are also interesting, in particular, the area distribution close to the percolation threshold and the characterization of how homogeneous these areas are. 

While exploring the properties of this simple model, we unveiled its connection with the backbone of the percolating cluster of ordinary percolation. 
Despite its simplifying assumptions, it nonetheless provides
a good basis for the study of the spatial extension of an epidemic spread.
In section II we describe the model and the quantities we considered to characterize its behavior. In section III we show and analyze our results, obtained through extensive numerical simulations. The discussion and some conclusions are presented in section IV.


\section{Model and Observables}

We consider the model introduced in Ref.~\cite{deOliveira21} where $N\leq L^2$ agents are 
initially distributed at random and without superposition on an $L \times L$ square lattice 
with periodic boundary conditions. 
As will be discussed at the end of this section, it is possible~\cite{NeZi00,NeZi01} to transform the results obtained with a constant, discrete $N$ to a continuous variable $p$. 
We choose an initial condition in which only one agent is infected (I) while the other $N-1$ are susceptible (S).
Starting from this initial configuration, at each Monte Carlo step (MCS), each agent sequentially chooses one of its nearest neighbor sites at random. 
If the later is vacant, the agent moves to that site. 
Otherwise, in case the agent is currently infected and the tossed neighbor is a susceptible, then the later also becomes infected with probability $p_{\small\rm inf}$. 
Each infected agent remains in the I state during $t_{\small\rm rec}$ time steps (kept constant), after which it becomes recovered (R) and unable to further propagate the disease to other agents. 
Notice that because of the excluded volume, recovered agents indeed help hindering the transmission of the disease. 
We here adopt $p_{\small\rm inf} = 0.5$ and $t_{\small\rm rec} = 20$ and discuss, in the conclusions, on the generality of the results. 
Notice that having $t_{\small\rm rec}>1$ makes the model non Markovian~\cite{MiBo13,KiRoVi15,FeCaTaLa19}. 
The above steps are repeated until no infected agent remains in the system, what defines the total time $T$ of the spreading process. 
The model also considers that the timescale of any demographic process is much longer than the contagion one, and the population is kept fixed. 
We also consider that displacements and contagion are local processes despite the complex
way humans are networked and the possibility of long range travels.

\begin{figure}[htb]
\includegraphics[width=0.6\columnwidth]{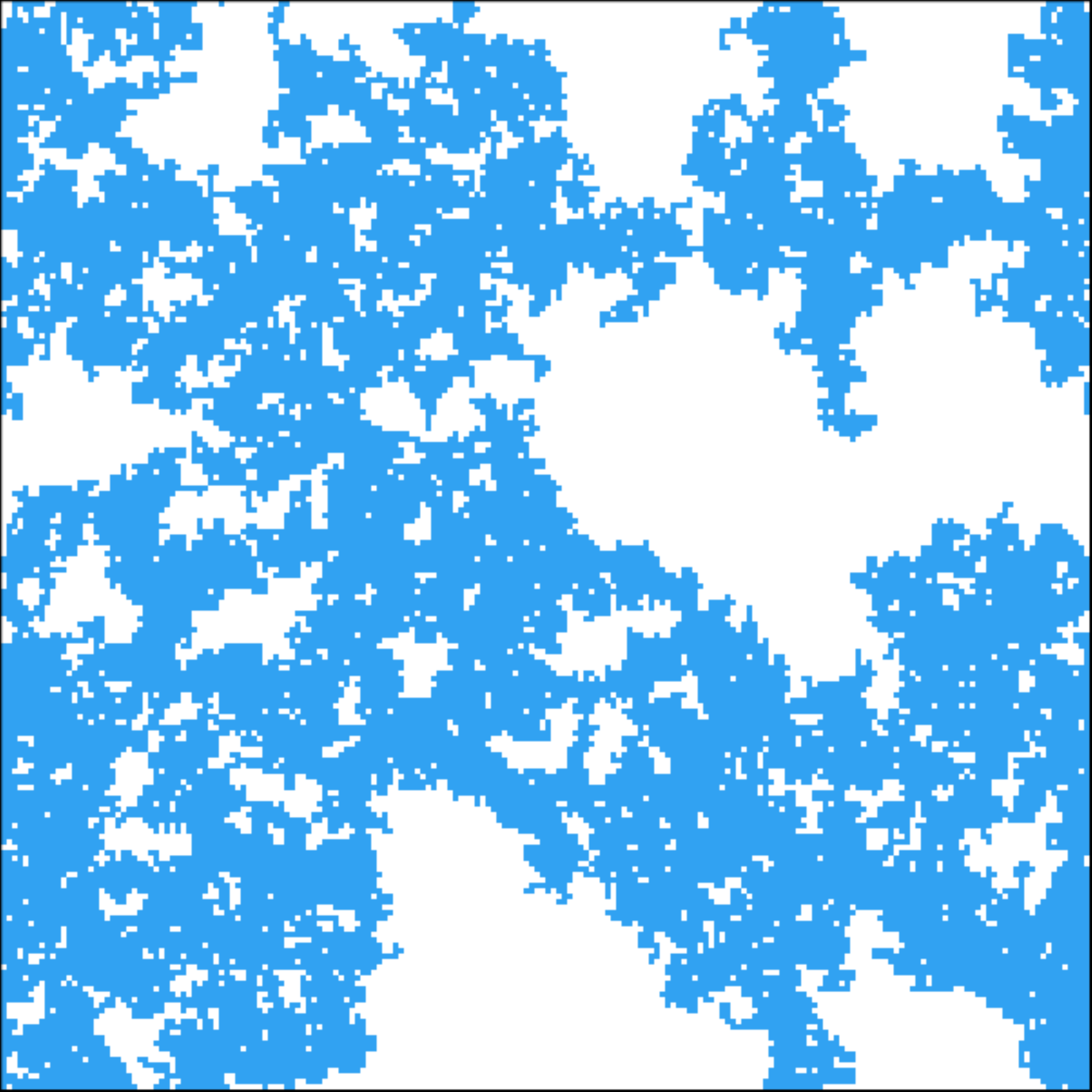}
\caption{Final configuration of the branching process for $L=200$ and $p=0.31$ (slightly above the percolation threshold) showing the single cluster formed by the sites visited by infected agents (blue). The remaining sites, on the other hand, are distributed among several clusters of varying sizes (white regions) that make the blue cluster fractal at the transition.}
\label{fig.snapshot}
\end{figure}

Here we focus on several properties of the many absorbing states attained at the end of the spreading process. Of primary interest is the single cluster, of size $M$, that includes all sites that have been visited by the infected agents.
An example of such a cluster is shown in Fig.~\ref{fig.snapshot} for $p$ slightly above the percolation threshold $p_c$ (to be determined later). 
The sample averaged fraction of sites belonging to the infected cluster is given by the strength $m = \langle M\rangle/L^2$, which acts as an order parameter for the propagation of the disease. 
It vanishes in the thermodynamic limit, $L \to \infty$, if the population density 
 is smaller than a critical value $p_c$, while it is finite for $p>p_c$.  
Thus, the epidemic threshold is associated with this cluster becoming macroscopic, i.e., occupying a finite
fraction of the system when $L\to\infty$.
The size fluctuations, $\chi = L^{-2}(\langle M^2\rangle-\langle M\rangle^2)$, and 
the associated Binder cumulant~\cite{Binder81}, $U = 1 - \langle M^4\rangle/3\langle M^2\rangle^2$, are computed as well. The latter has the interesting property of having null anomalous 
dimension, i.e., in the thermodynamic limit $U$ is a step function with two 
constant trays, one for $p<p_c$ (whose height depends on the nature of the transition)
and another one at $U = 2/3$ for $p>p_c$. 
The value that $U$ assumes exactly at $p_c$ is an indication of the order of
the transition. For continuous transitions this isolated value is intermediary 
between both trays. On the other hand, for discontinuous transitions 
the cumulant at $p_c$ assumes a non trivial value~\cite{ChLaBi86,LeKo91,SeSh05,MaFyGu14} below both trays.

Above $p_c$,  the outbreak cluster percolates in a finite system, i.e., it wraps around the system and touches the opposite boundaries. Besides the Binder cumulant, two other step functions with null anomalous dimension are useful in this context: the wrapping 
probability $W$ around the torus and the connection probability $Y$~\cite{Oliveira92,Oliveira94,FiMoOl94}.
Between two parallel lines a distance $L/2$ apart from each other, $Y$ is 1 if both sites belong to the
outbreak cluster and 0 otherwise. Since one can scan $L$ different pairs 
of parallel lines for each configuration ($L/2$ along each of the torus main 
directions), this observable can be computed with more precision than $W$.

Besides the single, connected cluster associated with the outbreak,  
the complementary set of the unvisited sites 
(Fig.~\ref{fig.snapshot}, white sites) is also of interest. Those domains, as they may become trapped inside
the outbreak cluster, may have many different sizes in the same sample. 
We measure the cluster size distribution close to the percolation threshold.
The number of distinct cluster sizes of a given configuration, $H$, indicates how heterogeneous  this distribution is and has been recently subject of study in different 
systems~\cite{NoLePa11,LeKiPa11,LvYaDe12,JoYiBaKi12,RoOlAr15,AzRoOlAr20,Mazzarisi2021}.  
In the limit $p\ll 1$, the unvisited sites form a large domain and
$H\simeq 1$. In the other limit, $p\simeq 1$, a few isolated holes of approximately unitary
size remain in the visited cluster and, being mostly of the same size, once again $H\simeq 1$. As $p$ moves to
intermediate values, $H$ increases because the structure of the outbreak cluster becomes more
complex and a set of interior holes develops, with different sizes. As a consequence of 
the distribution of cluster sizes being a power-law close to the threshold, $n(s)\sim s^{-\tau}$ ($\tau$ is the Fisher exponent), $H$ develops a peak that grows as $H_{\text{peak}}\sim L^{2/\tau}$.

Finally, following Refs.~\cite{NeZi00,NeZi01}, if all $L^2$ values of a given observable $X_N$ have been measured for a constant, discrete $N$, a transformation to a continuous variable $p$ is obtained by
\begin{equation}
X(p) = \sum_{N=1}^{L^2} \binom{L^2}{N} p^N (1-p)^{L^2-N} X_N .
\end{equation}
This procedure is equivalent to the traditional transformation from the microcanonical 
to the canonical ensemble. Notice that although in principle all the values
$N\in [1,L^2]$ should be considered in the above sum, because the coefficients of $X_N$ are
highly peaked, it is the neighboring region to the specific value of $N$  when $N/L^2 \sim p$ that  contributes the most. 
Of course, it is also possible to populate the initial system with a probability $p$ of occupying each site, the results being consistent. A large number of 
samples has to be considered, nonetheless, in order to achieve the desired precision. 
Indeed, for the averages shown here, no less than $10^5$ samples have been used. 
 
\section{Results}

\subsection{Outbreak Cluster}

\begin{figure}[hbt]
 \includegraphics[width=\columnwidth]{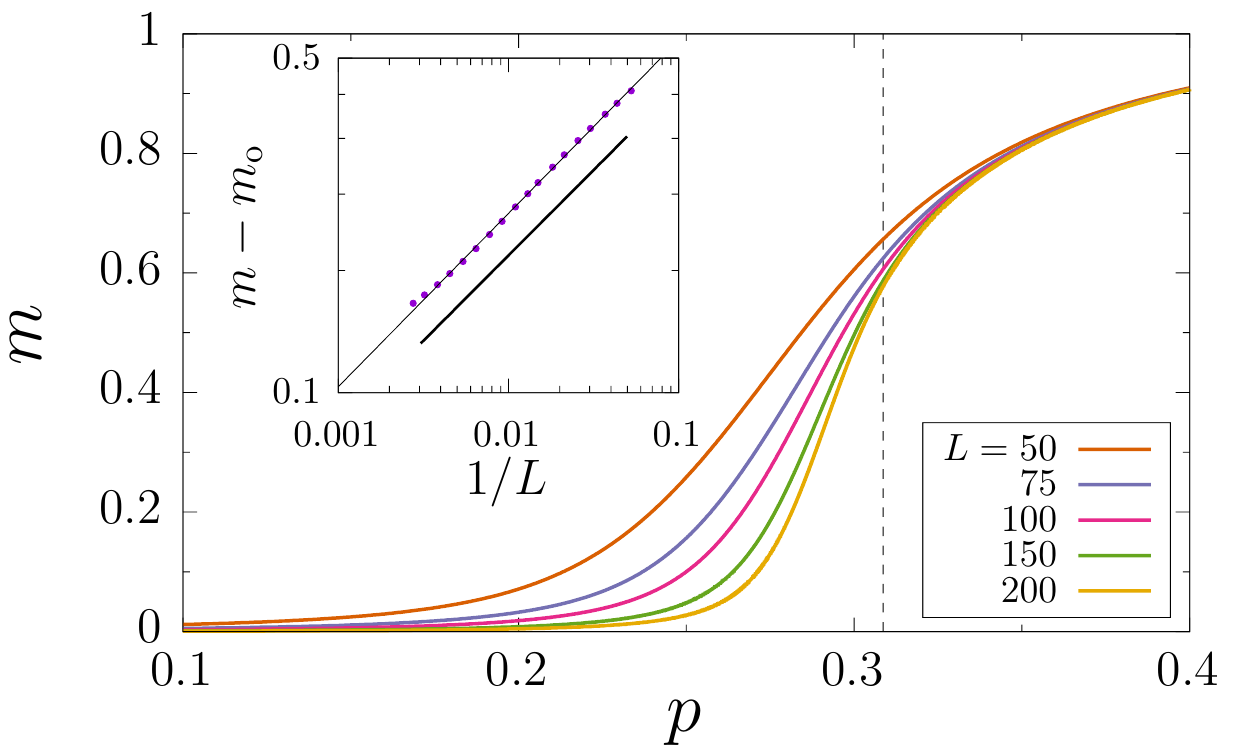}
\caption{The infected agents visit, on average, $\langle M\rangle$ sites during the outbreak, forming a single cluster that occupies a fraction $m\equiv \langle M\rangle/L^2$ of the system. There is a threshold $p_c$ (indicated by the vertical dashed line), separating a region ($p<p_c$) where this cluster is finite from another one, $p>p_c$, with a giant, percolating cluster. As the system size increases, the curves become steeper. Inset: Difference between the order parameter $m$ and the jump $m_{\rm o} \simeq 0.3070$ as a function of $1/L$, 
at the estimated critical point $p_c = 0.3086$. The best fit (thin solid line) gives $\beta/\nu\simeq 0.362$.  The thick solid line is $L^{2-d_B}$, where 
$d_B\simeq 1.643$ is the backbone fractal dimension (see text).}
\label{fig.m1}
\end{figure}

The set of evidences presented in this section points to a hybrid transition between a phase with a finite, non-percolating cluster of sites visited by the infected agents and another one with a giant, percolating cluster. 
In a hybrid, or mixed-order transition,  the order parameter has a finite jump at $p_c$.
Nonetheless, criticality remains after discounting the size of the jump from $m$, $m-m_{\rm o}\sim L^{-\beta/\nu}$, corresponding to a critical cluster of mass
\begin{equation}
M = m_{\rm o} L^2 + m_1 L^{d_F},
\end{equation}
where $d_F$, the cluster fractal dimension, obeys the hyperscaling relation $d_F=d-\beta/\nu$ with the exponent $\beta$ defined by  $m-m_ {\rm o}\sim (p-p_c)^{\beta}$.
Thus, the critical cluster has a compact component (first term) along with
a fractal part (second term). In standard second order transitions, $m_{\rm o}\to 0$ and the
compact region is missing. Discontinuous transitions miss the fractal term as $\beta\to 0$.
 In Fig.~\ref{fig.m1}, the order parameter $m$ is shown as a function of $p$ for different system sizes $L$ (the critical point, $p_c\simeq 0.3086$, that will be
 more precisely determined later, is shown as a vertical dashed line). 
 As $L$ increases, the curves become steeper. The behavior of $m-m_{\rm o}$ as a function of $1/L$ at the critical point $p_c$ is shown in the inset along with a power-law fit (thin solid line). From this, we get the exponent $\beta/\nu \simeq 0.362$ and the jump, $m_{\rm o} \simeq 0.3070$ (no sensible change is found in the exponent when using different values of this constant). For reasons that will be discussed later, we also show (thick solid line) that this behavior is consistent with the cluster having the same fractal dimension 
as the random percolation backbone, i.e., $L^{2-d_B}$, where $d_F=d_B\simeq 1.643$. The backbone is the subset of the critical cluster without the dangling ends, i.e., those sites that are not relevant for the transport properties through the cluster~\cite{StAh94}. 

\begin{figure}[hbt]
 \includegraphics[width=\columnwidth]{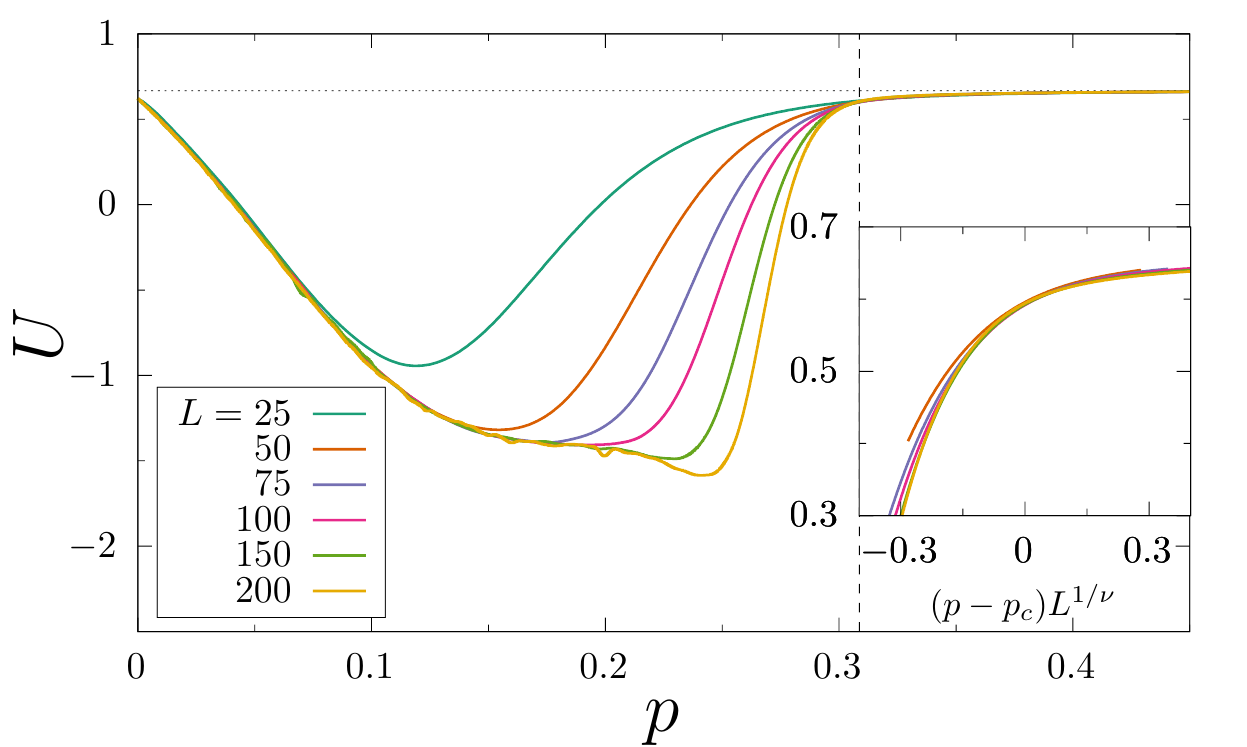}
\caption{The Binder cumulant for different system sizes. As $L$ increases, the minimum of $U$ moves to the right, approaching the transition point $p_c$. Inset: $U$ versus $(p-p_c)L^{1/\nu}$ showing that, for sufficiently large sizes, there is a very good collapse with $p_c\simeq 0.306$ and $\nu=2$.}
\label{fig.binder}
\end{figure}

Further evidence that the transition is not continuous is given by the Binder cumulant $U$, as shown in Fig.~\ref{fig.binder} (main panel) for several lattice
sizes. 
The behaviour is different from the typical one for continuous transitions. It shows a region of negative values with a minimum on the left of $p_c$. 
For small values of $p$, instead of forming a flat plateau at $U=2/3$, it converges to
a value below 2/3, probably due to the finite, increasing values of $M$ (indeed, even
for $p\to 0$, when $1 \leq M \leq t_{\small\rm rec}$, we get $U$ slightly below 2/3). 
Moreover, increasing $L$, the location of the minimum shifts towards the critical point $p_c$ and seems to converge to a non trivial, negative value. 
This is an indication that, being a zero anomalous dimension quantity, 
in the thermodynamic limit $U$ assumes an isolated value at $p_c$. 
The scaling behavior of $U$ is shown in the inset of Fig.~\ref{fig.binder}. 
In the critical region $U=f[(p-p_c)L^{1/\nu}]$ ($f$ is a universal function) and a very good collapse is obtained with $\nu=2$ and $p_c\simeq 0.306$ for the largest sizes.

\begin{figure}[!hbt]
 \includegraphics[width=\columnwidth]{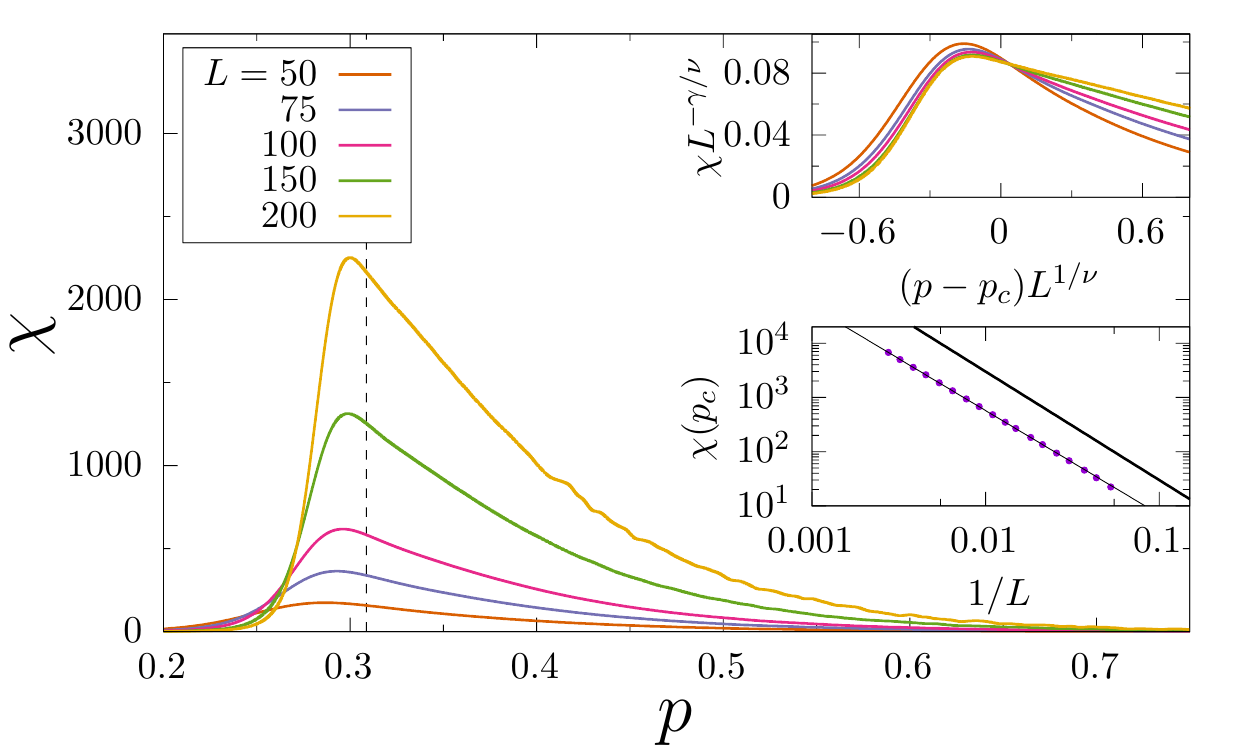}
\caption{Behavior of the susceptibility $\chi$ as a function of $p$ for different values of $L$, showing
the presence of a pronounced peak close to the transition. (Top inset) Data collapse of $\chi$ around the transition using $\gamma/\nu\simeq 1.91$, $\nu=2$ and $p_c = 0.3086$. Although finite size corrections appear for the smaller sizes, the two largest sizes are well superposed. (Bottom inset) 
Power-law increase of $\chi$ at $p=p_c$, $\chi(p_c)\sim L^{\gamma/\nu}$ where, from the fit (thin solid line), $\gamma/\nu\simeq 1.91$. The thick solid line shows the comparison with $\gamma/\nu=2$ (see text). In this inset, extra sizes were considered, from $L=19$ up to 363.}
\label{fig.chi}
\end{figure}

The fluctuations on the size of the outbreak cluster also have a non-trivial exponent, $\chi \sim |p-p_c|^{-\gamma}$.  
The main panel of  Fig.~\ref{fig.chi} shows that the susceptibility $\chi$, as a function of $p$, develops an increasing peak that moves toward $p_c$. 
In the bottom inset, we plot the height at $p_c\simeq 0.3086$, $\chi(p_c)$, versus $L^{-1}$ in a log-log scale. 
From the best fit (thin solid line) we obtain that its anomalous dimension is $\gamma/\nu \simeq 1.91$. 
The top inset shows the collapse using this value for $\gamma$ and $\nu=2$. 
Although deviations are present for the small sizes, for  sufficiently large systems (the largest two), the collapse is very good. 
The total time $T$ for all infected agents get removed also has a peak that moves toward $p_c$, but its height increases linearly, $T\sim L$ (not shown).
As will be shown in the next section, the distribution of cluster sizes for those sites {\em not} visited by the infected ones is a power-law at $p_c$, $n(s)\sim s^{-\tau}$, with an exponent that is clearly smaller than 2. 
It was argued in Ref.~\cite{Sheinman15} (see also Refs.~\cite{Sheinman15,ShShMa16,PrLe16,HuZiDe16}) that for $\tau<2$, $\gamma=1/\sigma$ and $\nu=1/\sigma d$. 
From these equations we obtain that $\gamma/\nu = d = 2$, i.e., $\gamma=4$, what is shown as a thick straight line in the bottom inset of Fig.~\ref{fig.chi} for comparison. Albeit our result is close to this value, there is still a clear difference.

\begin{figure}[!hbt]
\includegraphics[width=\columnwidth]{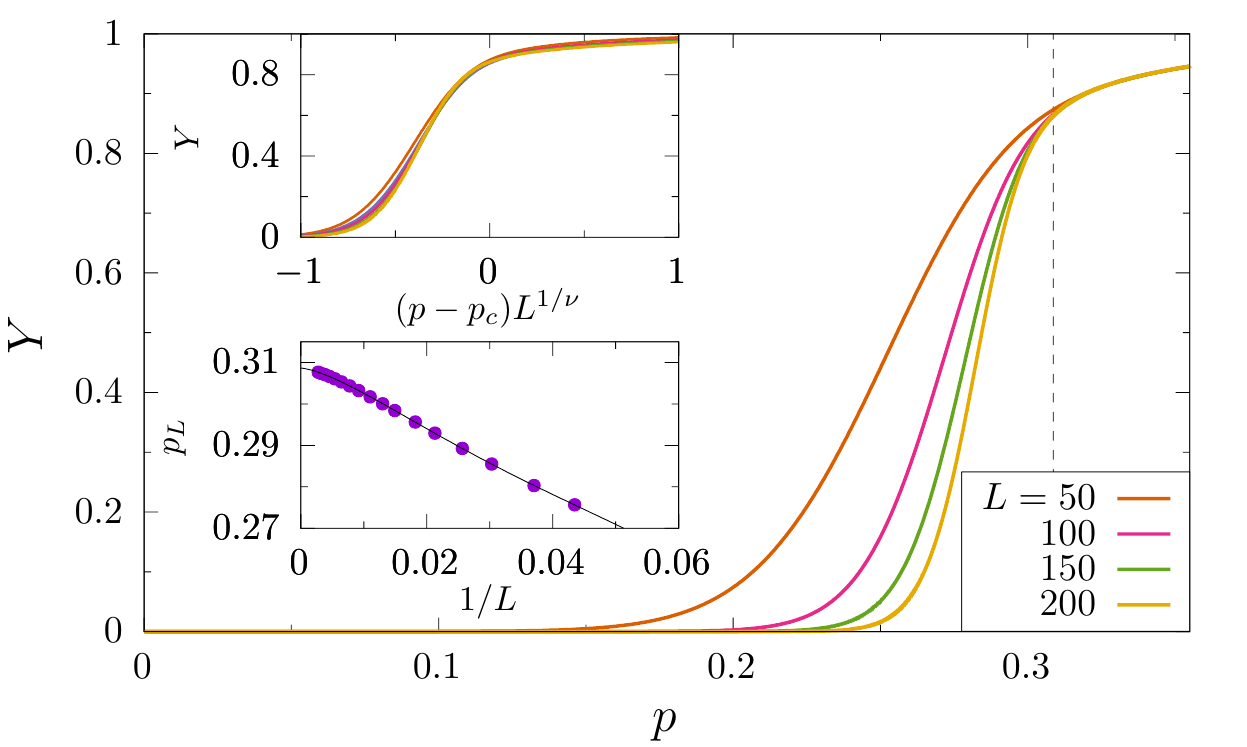}
\caption{The connection probability $Y(p)$ for several values of $L$. Top inset: data collapse onto a universal curve $g$ using $Y=g[(p-p_c)L^{1/\nu}]$.  Bottom inset: Corrections to finite size scaling. The black curve shows the fitting using Eq.~(\ref{eq.fit}) for the adopted value $\lambda = 0.7943$, which leads to $a_0 = 0$. The critical point is then
$p_c = 0.3086$. Extra sizes were considered, from $L=19$ up to 363.}
\label{fig.YL}
\end{figure}

The connection probability $Y(p)$ is shown in Fig.~\ref{fig.YL} for different system sizes.
Because of its null anomalous dimension, the vertical scale does 
not change and the scaling is given by $Y=g[(p-p_c)L^{1/\nu}]$, where $g(x)$ is a universal function. As for the Binder parameter, the scaling variable is $(p-p_c)L^{1/\nu}$ and, therefore, $1/\nu$ is the leading exponent for the scaling transformation along the horizontal axis.
In the top inset of Fig.~\ref{fig.YL} we observe that the collapse is very good with $\nu=2$ and $p_c\simeq 0.3086$. This precise value for $p_c$ will be determined in the sequence using the data for $Y(p)$. 

Following the method of Refs.~\cite{OlNoSt03,OlNoSt04}, one determines the point $p_L$ where $Y$ crosses a constant value $\lambda$, i.e., $Y(p_L) = \lambda$, for each $L$. 
Considering corrections to scaling up to the third order, this sequence of crossing points is fitted with the series expansion
\begin{equation}
p_L(\lambda) = p_c + \frac{1}{L^{1/\nu}} \sum_{n=0}^3 \frac{a_n(\lambda)}{L^n},
\label{eq.fit}
\end{equation}
with parameters $p_c$ and $a_n=a_n(\lambda)$. 
Since $\nu > 1$, the limit $1/L \to 0$ is problematic if $a_0\neq 0$. In this case, the curve $p_L$ versus $1/L$ will touch the vertical axis at $p_c$ as expected, but with a diverging derivative, what obviously weakens the accuracy in determining $p_c$. However, carefully choosing the value of $\lambda$, it is possible to obtain $a_0 \simeq 0$ and a null slope where $p_L$ touches the vertical axis. 
This is the case of the so-called Pinson number adopted for $W$ in the percolation problem~\cite{NeZi00}, whose value is exactly known from conformal invariance arguments, and corresponds to the value that the step function assumes at the isolated point $p_c$ in the thermodynamic limit. 
In our case, one needs to find the proper $\lambda$ value by the fitting procedure itself and approaching $a_0 = 0$, as in Refs.~\cite{OlNoSt03,OlNoSt04}.
Fitting Eq.~(\ref{eq.fit}) with $\nu=2$, one can obtain the coefficients of the corrections to scaling and evaluate, with good precision, the location of the critical point. Indeed, $\lambda \simeq  0.7943$ tunes $a_0 \simeq 0$, leading to $p_c \simeq 0.3086$ as shown in the bottom inset of Fig.~\ref{fig.YL}. The estimated uncertainty is located at the last digit which does not sensibly change even when discarding the larger lattice sizes.




\subsection{Uninfected Regions}

\begin{figure}[htb]
\includegraphics[width=\columnwidth]{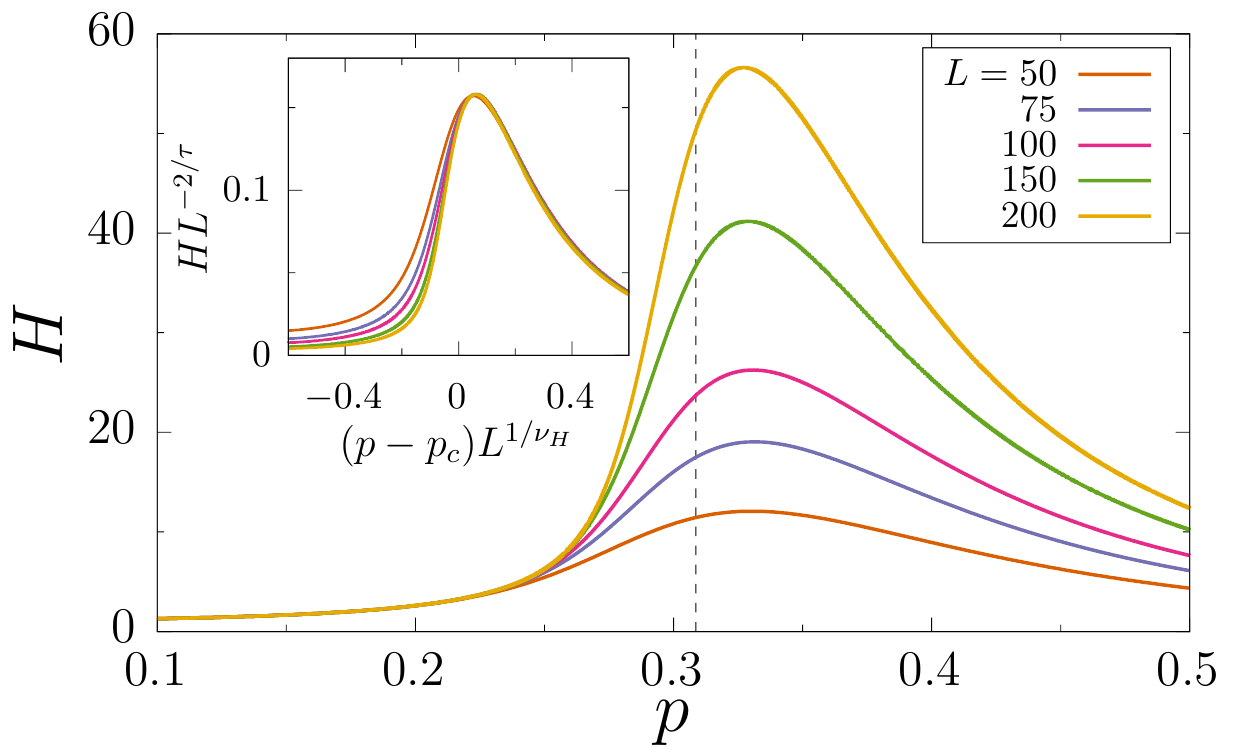}
\caption{(Main panel) Heterogeneity $H$ as a function of the occupation $p$ for several system sizes.
As in the ordinary percolation, for sufficiently large $L$, $H$ has a peak that approaches the threshold and grows as $H^*\sim L^{1.11}$. (Inset) Collapse of $H$ with $\tau\simeq 1.8$, $p_c\simeq 0.308$ and $\nu_H=(\tau-1)/\nu\tau$ with $\nu\simeq 2$. We do not consider the smallest sizes because of the strong finite size effects.}
\label{fig.H}
\end{figure}

We now consider the geometrical properties of the unvisited sites close to the percolation threshold, occupying the space left by the outbreak cluster. 
Although the latter is a single cluster, the sites never visited by an infected agent may form many disjoint clusters, as seen in Fig.~\ref{fig.snapshot}.
For each sampled configuration, the number of different sizes with at least one cluster present gives, once averaged, the size diversity (or heterogeneity) $H$~\cite{NoLePa11,LeKiPa11,LvYaDe12,JoYiBaKi12,RoOlAr15,AzRoOlAr20,Mazzarisi2021}. 
The results are shown in Fig.~\ref{fig.H} for several values of $L$. 
As expected, it presents a growing peak, of height $H^*$, at a concentration $p^*$ close to the threshold. 
Similarly to the ordinary percolation~\cite{NoLePa11}, these peaks are located in the region where the outbreak cluster percolates while the largest uninfected cluster is still building up (in our case, $p>p_c$). 
The position of the peak moves to the right if one only considers the smaller sizes.
However, for intermediate sizes it starts moving toward the previously estimated threshold $p_c$.
This is an indication that the position of the peak of $H$, for the sizes we are able to simulate, does not provide a reliable extrapolation for the threshold. 
This pre-scaling, strong finite-size effect prevents a more precise estimate of the Fisher exponent $\tau$, that is related with the height of the peak and with the power-law tail of the size distribution at the threshold.

Fitting the height of the peaks in Fig.~\ref{fig.H}, we obtain that $H^*\sim L^{1.11}$. Considering
that $H^* \sim L^{d/\tau}$~\cite{NoLePa11,RoOlAr15}, it implies that the Fisher exponent is smaller than 2, $\tau\simeq 1.8$. This indeed provides a good collapse
for the height, as shown in the inset of Fig.~\ref{fig.H}.  All peaks
can be centered with $p_c\simeq 0.308$, a value that is consistent with the previous
estimates. As discussed in Refs.~\cite{NoLePa11,RoOlAr15}, the width of the critical region
scales with $L^{1/\nu_H}$ instead of $L^{1/\nu}$, 
where $\nu_H = \tau\nu/(\tau-1)$.
With $\nu\simeq 2$, as can be seen in the inset of Fig.~\ref{fig.H}, the collapse is excellent.

\begin{figure}[htb]
\includegraphics[width=\columnwidth]{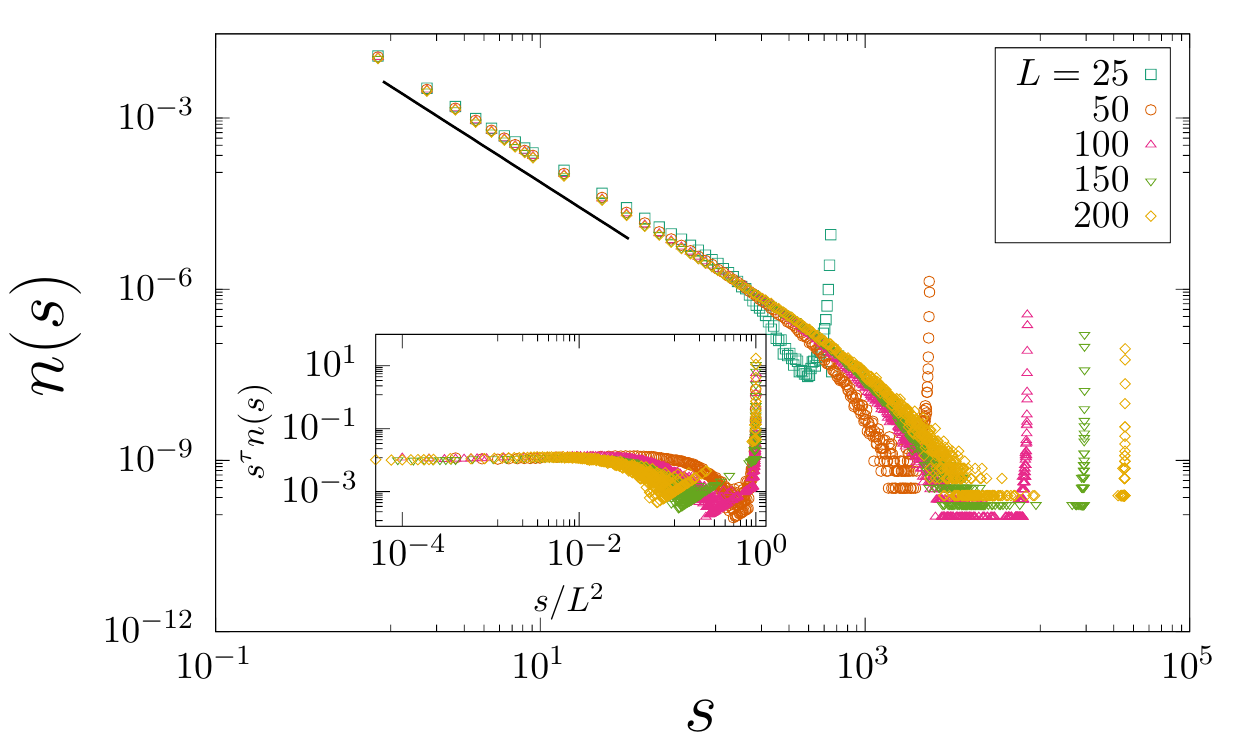}
\caption{Cluster size distribution for the unvisited sites measured at the peak of $H$ (see Fig.~\ref{fig.H}) for several system sizes. For comparison, we show (solid line) the $s^{-1.8}$ behavior of $H^*$, that is also consistent with the distribution of holes in the percolation backbone~\cite{HuZiDe16}. Inset: the flat region corresponds to the power-law behavior of $n(s)$.}
\label{fig.dist}
\end{figure}

Fig.~\ref{fig.dist} shows the size distribution $n(s)$ for the uninfected regions when the number of
particles (not $p$) is kept constant and corresponds to the point where $H$ has a peak, i.e., $N\simeq p^*L^2$. 
Since $p^*>p_c$, this peak occurs in the region where the outbreak cluster has a large probability of percolating. Thus, the complementary set of unvisited sites can percolate in this region only when the 
outbreak is halted very early, in the first steps of the dynamics. Once this happens,
the percolating cluster will occupy most of the lattice and contribute to the peak located close to $s \sim L^2$, in the tail of the distribution. The initial part of the distribution is a power-law,
$n(s)\sim s^{-\tau}$ whose width increases with $L$ because $p^*$ slowly approaches $p_c$. The exponent is consistent with our previous estimate obtained from the height of $H$ at $p^*$, $\tau\simeq 1.8$, as can be seen in the figure (solid line). This value is compatible with the exponent of the size distribution of holes in the backbone of the random site percolation cluster~\cite{HuZiDe16}, with the no-enclave percolation (NEP) model~\cite{Sheinman15} and with the clusters formed by sites not
visited by a random walk~\cite{FeKa21}.  Moreover, for large values of $s$, 
there is a peak corresponding to a percolating unvisited cluster. In the NEP model, clusters that are fully surrounded by larger clusters are absorbed into the latter. A similar effect occurs in our model while the
many infected agents are randomly walking and visiting most of the sites around it. Both mechanisms decrease
the number of enclaves in the large cluster.

\section{Conclusions}

We studied, from a statistical mechanics point of view, the equilibrium properties of a simple model for a disease outbreak. 
In a population of $N$ mobile agents on a square lattice, a single initially infected agent may transmit the disease to nearest neighbors. 
After a given period of time, $t_{\small\rm rec}$, an infected agent becomes unable 
to further transmit the disease and gets recovered. 
We define the outbreak cluster as the set of connected sites visited by the infected agents starting from the patient zero. 
For low densities, there is a large probability of the outbreak being halted at the early steps of the dynamics. 
In this case, the outbreak cluster would contain only a few sites. 
As $N$ increases, the outbreak cluster also gets larger and, above a given threshold, it percolates through the lattice. 
Since infected agents become recovered after $t_{\small\rm rec}$ steps, the interior of the outbreak cluster may not get completely filled and many holes of different sizes may be present. 
We focus here on the geometrical properties of the single outbreak cluster and of the set of holes. 
Since there is not a simple way to make our algorithm incremental, we cannot use the 
full power of the Newman-Ziff algorithm~\cite{NeZi00,NeZi01}. 
Nonetheless, the measures are obtained with a constant $N$ and then reweighted following their procedure. 
In this way, our control variable is $p$, a continuous parameter equivalent to the density, or the probability of having an agent in each site. 
The results show a hybrid percolation transition at $p_c\simeq 0.30865$ (for the set of parameters chosen here), along with estimates for the critical exponents that strongly indicate that the generated outbreak cluster is in the same universality class of the random percolation backbone. 

The backbone consists of a set of blobs connected by single links~\cite{HeSt84}. 
In the model we considered here, when a site get infected, the next sites visited by its random-walk are included in the outbreak cluster until it gets removed. 
Each one of these regions is similar to a blob. 
When, through contact, a susceptible agent get infected, it starts a new blob. 
The ensemble of blobs compose both the backbone and the outbreak cluster. 
The set of sites belonging to the backbone is usually identified among the larger set of the percolation cluster. 
Our model also provides a way of building a cluster whose critical properties are the same as the the backbone of ordinary percolation, although a few dangling ends may remain.

Close to the threshold, the cluster size distribution of the holes is a power-law whose exponent $\tau$ is smaller than 2.
Our result is compatible with the value obtained from the holes in the percolation backbone, whose $\tau$ is given by the Mandelbrot hyperscaling relation
\begin{equation}
\tau = 1 + \frac{d_B}{2} \simeq 1.822,
\end{equation}
where  $d_B\simeq 1.643$ is the fractal dimension of the backbone~\cite{HeSt84}. 
Once $d_B$ (or $\tau$) is known, other exponents can be determined. 
For example, $\beta/\nu = d -d_B \simeq 0.357$ and $\gamma/\nu=d=2$.
These exponents are very close to the values we obtained for our model, with the exception of
$\gamma$ where we obtained $\gamma \simeq 1.91$ instead of 2.

Whether the general conclusions drawn from the chosen parameters change with different sets of values, generating a richer phase diagram, is still to be verified.
For example, if $t_{\small\rm rec}$, the time an agent remains infectious, were larger, each infected agent would have more time to diffuse and fill more holes in the outbreak cluster. 
However, the whole cluster would grow larger as more agents will get infected. 
The question is whether the outbreak cluster will get more compact or, for sufficiently large systems, its fractal dimension will remain unchanged. 
A similar question applies for a smaller value of $t_{\small\rm rec}$. In this case, the contribution of each infected agent to the outbreak cluster will be less compact, perhaps approaching the self-avoiding random walk case.
In addition, it would be important to consider larger system sizes to confirm and extend the above results. 
Although constrained by the excluded volume condition that prevents more than one
individual in the same site, the mobility of the agents induce some local shuffling
and an effective longer interaction. 
Thus, it would be interesting to approach this problem analytically and check how well a mean field approximation would describe the results presented here. 
Also, the model considered here can be studied on a non-regular network in order to check whether the transition remains of mixed-order~\cite{Watts02}.

\begin{acknowledgments}
Work partially supported by the Brazilian agencies FAPERJ, Conselho Nacional de Desenvolvimento Cientí\-fi\-co e Tecnológico (CNPq), and CAPES. 
\end{acknowledgments}

\bibliographystyle{apsrev}   

\end{document}